\documentclass[twocolumn,prd,amssymb,amsmath,preprintnumbers,floatfix,aps,nofootinbib]{revtex4-1}
\pdfoutput=1
\usepackage{dcolumn,graphicx,bm,psfrag}
\usepackage[colorlinks=true,citecolor=blue,urlcolor=black,linktocpage=true,
linkcolor=blue]{hyperref}
\usepackage[dvipsnames]{xcolor}

\newcommand{\be}{\begin{equation}}
\newcommand{\ee}{\end{equation}}
\newcommand{\bear}{\begin{eqnarray}}
\newcommand{\eear}{\end{eqnarray}}
\newcommand{\ba}{\begin{array}}
\newcommand{\ea}{\end{array}}
\setcitestyle{numbers}
\usepackage{eurosym}

\begin{document}
\preprint{TUM-1188/19} 
\title{Current and future perspectives of positronium and muonium spectroscopy \\ as dark sectors probe}

\author{Claudia Frugiuele}

\affiliation{Department of Particle Physics and Astrophysics, Weizmann Institute of Science, Rehovot 7610001, Israel}

\author{Jes\'{u}s P\'{e}rez-R\'{i}os}

\affiliation{Fritz-Haber-Institut der Max-Planck-Gesellschaft, Faradayweg 4-6, D-14195 Berlin, Germany }

\author{Clara Peset}
\affiliation{Physik Department T31, James-Franck-Stra\ss e 1, Technische Universit\"at M\"unchen,
85748 Garching, Germany }
\date{\today}

\begin{abstract}
Positronium and Muonium are purely leptonic atoms and hence free of an internal sub-structure.
This qualifies them as potentially well suited systems to probe the existence of physics beyond the Standard Model.  We hence carry out a comprehensive study of the sensitivity of current Positronium and Muonium precision spectroscopy to several  new physics scenarios. By taking properly into account existing experimental and astrophysical probes, we define clear experimental targets to probe new physics via precise spectroscopy. For Positronium we find that, in order for the spectroscopy bounds to reach a sensitivity comparable to the electron gyromagnetic factor, an improvement of roughly five orders of magnitude from state-of-the-art precision is required, which would be a challenge based on current technology. More promising  is instead the potential reach of Muonium spectroscopy: in the next few years  experiments like Mu-MASS at PSI will probe new regions of the parameter space testing the existence of medium/short range (MeV and above) spin-dependent and spin-independent dark forces between electrons and muons.


\end{abstract}

\maketitle

\section{Introduction}
The Standard Model (SM)  provides a remarkably successful description of particle physics.  However, there are several reasons why new physics  beyond the Standard Model (BSM) should exist such as the existence of Dark Matter (DM), a yet-unidentified form of matter which permeates the whole Universe. Another unsolved puzzle stems from the origin of neutrino masses and mixing, which could be the portal to a richer sector. Scenarios aimed at explaining BSM puzzles  generically predict the existence of dark sectors which consist of light sub-GeV particles very weakly coupled to the visible sector \cite{quinn,kaplan,relaxionP}. These particles might be copiously produced at fixed-target experiments and low energy colliders (so-called \textit{intensity frontier})  or searched for in precision measurements experiments (so-called \textit{precision frontier}).  A solid experimental investigation requires the coexistence of  both intensity and precision frontier probes. On the one hand, precision frontier probes have the advantage of not depending on model details (e.g the decay mode of such dark particles) thus simultaneously exploring a large set of scenarios. On the other hand, the program at the intensity frontier can provide crucial information on dark sector properties. Dark sectors particles induce a new feeble  intermediate/long-range force within the visible sector itself, which could leave a footprint in precise atomic spectroscopy measurements.

 Leptonic atoms represent a solid playground to investigate the existence of such dark forces since they are free of nuclear effects. 
 However, these systems are experimentally very challenging, due to their short lifetimes. Among the different leptonic atoms, Muonium (Mu=$\mu^{+}e^{-}$) and Positronium (Ps=$e^{+}e^{-}$) are promising systems to consider in the quest of BSM physics~\cite{KNP1,KNP2,mckeenK}. The formation of Mu and its lifetime is limited by the unstable character of $\mu^{+}$, which in 2.196 $\mu$s decays into a $e^{+}$ through the electro-weak interaction, which makes Mu a tricky system to deal with in the laboratory. Nevertheless, undertaking high precision spectroscopy of Mu will help to test QED as well as to resolve the electron to muon mass ratio up to 1ppb  \cite{Crivelli:2018vfe,MUSEUM}. On the contrary, formation of Ps is easily achieved in the laboratory since both constituents are stable, although it shows an average lifetime of 142 ns~\cite{Rich:1981nc}, dictated by the $e^{+} - e^{-}$ annihilation rate weighted by the Ps wavefunction at the origin. Ps and Mu have been already proposed to test QED and BSM physics~\cite{JPR2018,Lamm,Ozeri,helium,KNP1,KNP2,mckeenK}.

In this paper we present a comprehensive study of the perspectives to hunt for dark sectors with Ps and Mu spectroscopy and we evaluate which measurements could be more suited to give possible interesting results in the future. To this purpose, we consider state-of-the-art measurements, commenting on their possible improvements and the perspectives for new measurements.
 One of the goals is investigating whether with novel measurements, it would be possible to reach comparable sensitivity to the one of the electron and muon gyromagnetic factors $a_e$ \cite{g2e,giudice2} and $a_{\mu}$ \cite{g2m}. 
We compare the sensitivity of spectroscopy to new physics only with other experimental probes at the precision frontier as the leptonic gyromagnetic factor. We do not consider probes at the intensity frontier since these are more model dependent and depend on the decay and production modes, see \cite{Alexander:2016aln} for a review.

\section{Spin-independent forces}
A spin-independent dark force between electron-positron or antimuon, mediated either by a new scalar (e.g. \cite{relaxionP,kaplan}) or a new vector gauge boson (if this corresponds to an anomalous gauge coupling, other strong constraints must be taken into account as discussed in \cite{Dobrescu:2014fca,Dror:2017nsg}), gives rise to a Yukawa-like attractive potential:
\begin{align}
	\label{eq:Ven}
	V^{ij}_{SI}(r) =  -\frac{g_i g_j }{4\pi}\frac{e^{-m_\phi r}}{r} \, ,
\end{align}
where $g_i$ is the dimensionless coupling constant to the lepton $i,j$ and $m_{\phi}$ represents the mass of the scalar/vector. Eq.~(\ref{eq:Ven}) leads to a modification of the atomic energy levels and hence to the frequency shift for a given transition $ a \to b$. An analytical expression of the energy levels produced by Eq.~(\ref{eq:Ven}) for general quantum numbers can be found in \cite{Claudia2018}.
Consider a measured transition corresponding to an experimental accuracy $\Delta E_{a\rightarrow  b}^{\rm exp} $ and a theoretical one $\Delta E_{a \rightarrow b}^{\rm theo}$.
Due to the agreement between theory and experiment we know that:
\be
|\Delta E_{a \rightarrow b}^{\rm BSM}| < | \Delta E_{a\rightarrow  b}^{\rm exp} -\Delta E_{a \rightarrow b}^{\rm theo} |   \lesssim2 \sigma_{\rm max},
\ee
where $ \sigma_{\rm max}$ is the biggest source error. This could come either from the experimental measurement or via the theoretical prediction, that is:
\be
\sigma_{\rm max}  \equiv \rm Max( \sigma_{\rm exp}, \sigma_{\rm theo}).
\ee
\subsection{Positronium}
\paragraph{ $1S - 2S$ transition }
 A new spin-independent dark force between electrons could be probed considering the $1S - 2S$ transition.
The current theoretical prediction is:
\be
(E(2^3 S_1)-E(1^3S_1))^{\text{th}} _{\text{Ps}} =1233607222.13(58) \;  \rm MHz,\label{PsLS}
\ee
where $E(n^{2s+1}L_J)$ denotes the energy of the electronic state within the parenthesis. Eq.~(\ref{PsLS}) includes all relativistic corrections to the tree level up to $\mathcal{O}(m_e\alpha^6)$ and the leading logarithms of $\mathcal{O}(m_e\alpha^7\ln^2\alpha)$, $\mathcal{O}(m_e\alpha^8\ln^3\alpha)$ computed in \cite{PhysRevA.59.4316,Manohar:2000rz}. The error is estimated as one half of the last two contributions. On the other hand, the best measurement is \cite{Ps1s2s} \be
(E(2^3 S_1)-E(1^3S_1))^{\text{exp}} _{\text{Ps}} =1233607216.4(3.2) \; \rm MHz.\label{PsLSexp}
\ee

Hence, considering current experimental precision BSM physics can be tested at  MHz level.
The consequent constraint on new physics was previously pointed out in Ref.~\cite{helium} where it was remarked that the current state-of-the-art in Ps reaches a comparable sensitivity to $a_e$ in the massless limit, $ m_{\phi} < 1/a_{0,e}$.
However, in this region of the parameter space, the electron coupling is severely constrained by astrophysics, i.e., by stellar cooling effects induced by new light degrees of freedom \cite{stellarLasenby}:
\be
(g_e)_{\rm astro} \lesssim 10^{-14},
\label{astro}
\ee
which applies to $ m_{\phi} \lesssim 300$ keV. 
This bound is so strong that it is difficult to imagine the possibility of testing via Ps precise spectroscopy relevant regions of the parameter space for new physics, even considering novel measurements or improvements of the existing ones (both on the theory and experimental side). 
A possible loophole is represented by models where very light bosons  which exhibit screening effects  such as the chameleon \cite{atom1,Joyce:2014kja}, thus evading  astrophysical constraints.
\par
The region of the parameter space where atomic spectroscopy could play a crucial role in testing new physics is instead the heavier mass region.
\begin{figure}[t]
\begin{center}
\includegraphics[scale=0.6]{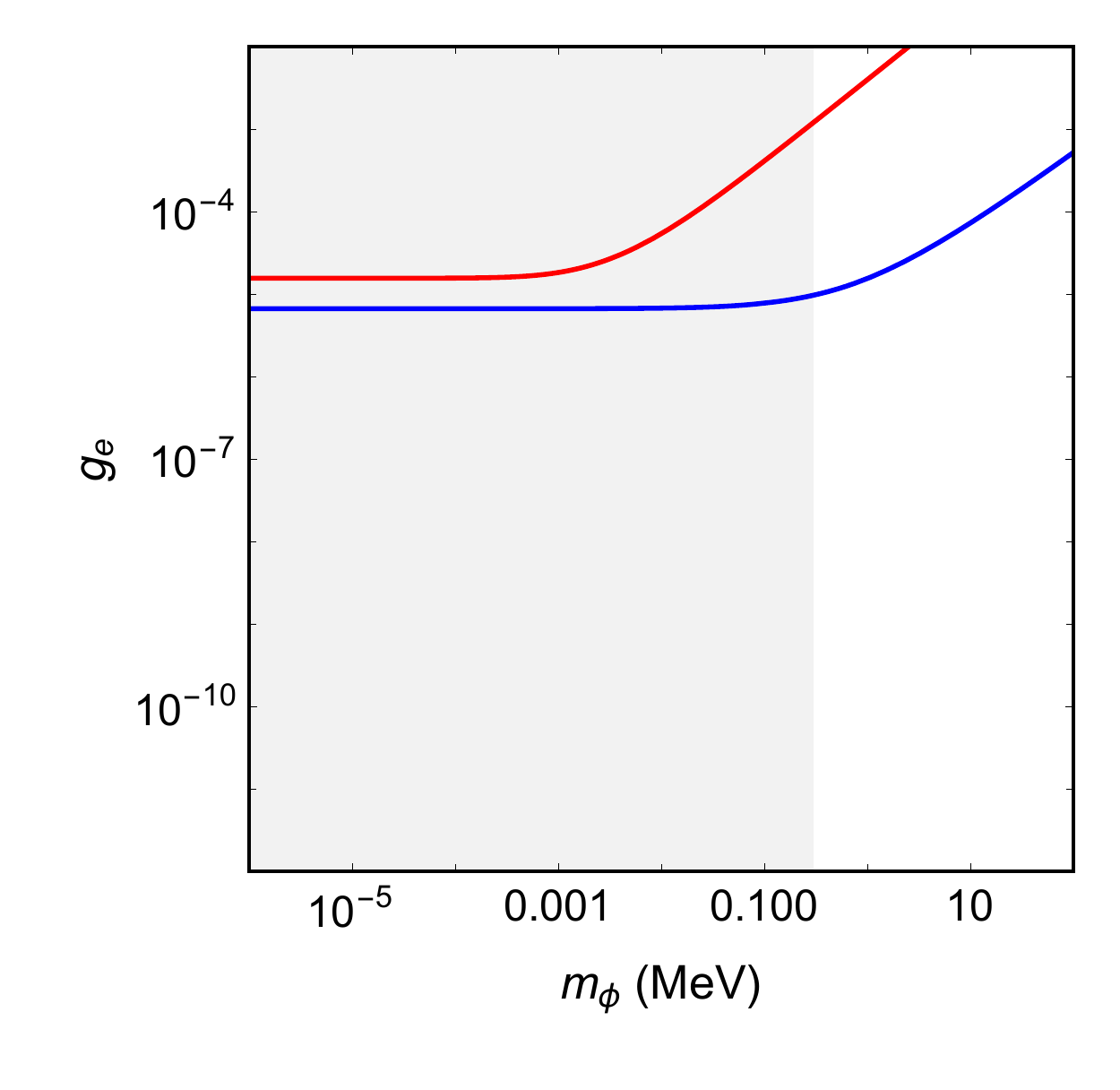}
\caption{Constraint on the dimensionless coupling  $g_e$ as a function of the scalar/vector mass. The blue curve represents the bound coming from the measurement of the electron gyromagnetic factor $a_e$ \cite{g2e,giudice2}, while the red curve is the current bound extracted from the Ps $1S-2S$ transition \cite{helium,Ps1s2s}. The gray region is excluded by astrophysics (i.e. stellar cooling) \cite{stellarLasenby}.}
\label{figure1}
\end{center}
\end{figure}
However, for $m_{\phi} >> m_e$, the constraint from $ a_e $ is significantly stronger than the Ps current reach as it is clearly shown in Fig.~\ref{figure1}.
The question we want to address regards the feasibility of overcoming the precision of $a_e$ with Ps spectroscopy.
Let us note that achieving this would have several potential advantages: atomic observables are sensitive to new physics at tree level while  $a_e$  starts at  loop level~\cite{giudice2} and hence is more prone to cancellation against additional contributions from other states present in a complete model. 
  \paragraph{ Rydberg transitions}
 The lifetime of Ps Rydberg states (states with large principal quantum number $n$) is not limited by the electron-positron annihilation, owing to its scaling as $n^{3}$. Thus, Rydberg Ps states can be seen as regular Rydberg atoms, which can be easily manipulated through external electric and magnetic fields. This has fueled the community to pursue exciting experiments related to the deexcitation of Rydberg states of Ps with $n\lesssim 30$~\cite{ref1,ref2,ref3}. High precision spectroscopy of Rydberg states of Ps could also offer new interesting possibilities to probe new physics. 
 \par
\begin{figure}[t]
\begin{center}
\includegraphics[scale=0.6]{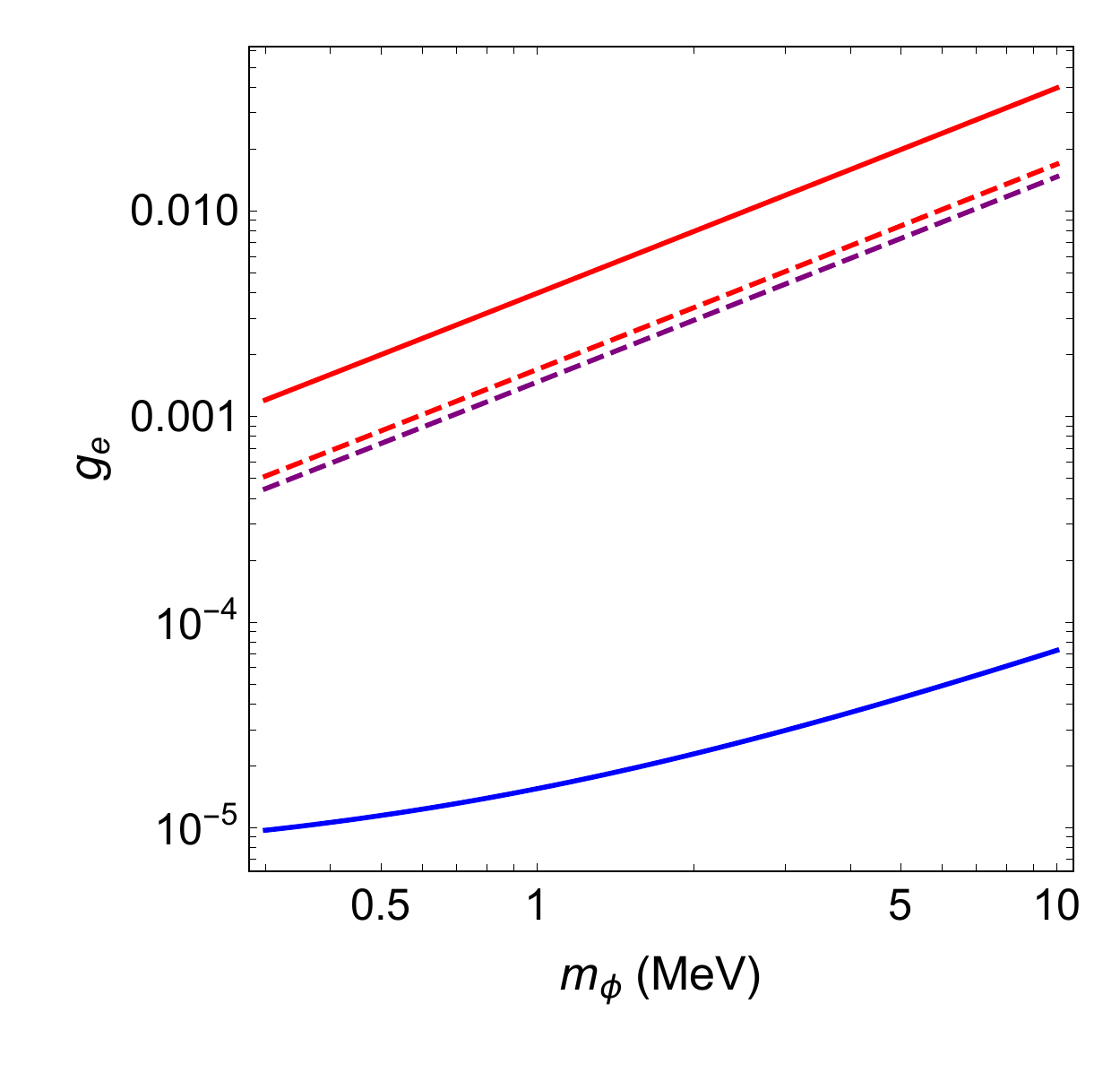}
\caption{Constraint on the dimensionless coupling  $g_e$ as a function of the scalar/vector mass. As in Fig.~\ref{figure1}, the blue curve represents the bound coming from the measurement of the electron gyromagnetic factor $a_e$ \cite{g2e,giudice2}, while the red curve is the current bound extracted from the Ps $1S-2S$ transition \cite{helium,Ps1s2s}.
The dashed red curve is the projected sensitivity assuming that the experimental precision will match the theoretical one in Eq.~(\ref{PsLS}). The dashed purple curve is the sensitivity of Rydberg transitions \cite{ref1,ref2,ref3} assuming a 500 kHz experimental and theoretical precision.}
\label{merda}
\end{center}
\end{figure}

 In Fig.~\ref{merda} we consider only the most relevant region of the parameter space (where astrophysical bounds do not apply, i.e. $m_{\phi} \gg 300$ keV)  and we compare the current  $a_e$ constraint both to the current reach for the Ps $1S-2S$ transition (red line) and to an estimation of Ps sensitivity to new physics based on a hypothetical novel measurement where the experimental precision could match the current theoretical precision in Eq.~(\ref{PsLS}) (red dashed line), for instance the planned measurement at $ 5 \times 10^{-10}$  at  ETHZ \cite{Cooke:2015gha}. 
 Furthermore, we evaluated the sensitivity of Rydberg spectroscopy for $1S-20S$ transition (purple-dashed line) to spin-independent forces assuming a theoretical and experimental precision of 500 kHz.  We have checked that the new physics sensitivity does not depend on the large principal quantum number $n \gg 1$, since only the short-range tail of the wave function affects the shifts of the levels involved.  Therefore, the sensitivity of Rydberg transition is comparable of that of the $1S-2S$ transition.
 Fig.~\ref{merda} clearly shows that this would not be sufficient to reach unexplored regions of the parameter space: orders of magnitude (precision down to 10 Hz) improvement is necessary to reach a competitive sensitivity.
 This level of precision is futuristic based on the current laser technology.

\subsection{Muonium}
Mu spectroscopy offers the possibility to probe the existence of new light degrees of freedom coupled to both electrons and muons.
This is particularly interesting for the possible relation of leptophillic spin-independent new forces  to the muon $g-2$ anomaly \cite{PhysRevD.80.095002}. Hence, investigating the perspectives of Mu spectroscopy to spin-independent forces has the potential to highlight relevant regions of the parameter space for such a longstanding puzzle. 
\paragraph{ $1S - 2S$ transition }

The current experimental measurement of the $1S - 2S$ transition is \cite{muonium1s2sEXP}
\be
(E(2 S_{1/2})-E(1S_{1/2}))^{\text{exp}} _{ \rm Mu} = 2 455 528 941.0(9.8)  \; \rm MHz,\label{Mu1s2sexp}
\ee
where now $E(n L_{J_e})$ denotes the energy of the spin-averaged muonium state within the parenthesis.
In the next few years, a new planned experiment at PSI, Mu-MASS \cite{Crivelli:2018vfe},  will improve the experimental precision down to the kHz level.

On the theoretical side, the muonium energy levels have been computed completely up to $\mathcal{O}(m_\mu\alpha^5)$ \cite{Peset:2015zga} and the leading logarithmic correction $\mathcal{O}(m_\mu\alpha^6\ln\alpha)$ \cite{1402-4896-1993-T46-040}.  The $1S- 2S$ transition has reached however $\mathcal{O}(m_\mu\alpha^7)$ \cite{Eides:2000xc} and so the QED error should be estimated by the $\mathcal{O}(m_\mu\alpha^8\ln^3\alpha)$ term, which would give $\sim 10$ kHz. However, the main source of uncertainty is not the QED computation but the value of the muon mass. The best value for the muon mass gives an uncertainty $\sim 0.3$ MHz, but this muon mass relies on the measurement of $1S- 2S$ and hyperfine splittings in muonium and so we cannot use it as an independent input of our theoretical estimate if we want to use it to set bounds on new physics. 
Therefore, we chose to consider the measurement of the muon mass determined from the study of Breit-Rabi magnetic sub-levels of the Mu ground state in an external magnetic field \cite{ osti_295569}, which would be unaffected by the new scalar particle. This gives rise to the theoretical prediction: 
   \be
(E(2 S_{1/2})-E(1S_{1/2}))^{\text{th}} _{\rm Mu} = 2 455 528 935.8(1.4) \;  \rm MHz.\label{Mu1s2stheo}
\ee

  \begin{figure}[t]
\begin{center}
\includegraphics[scale=0.6]{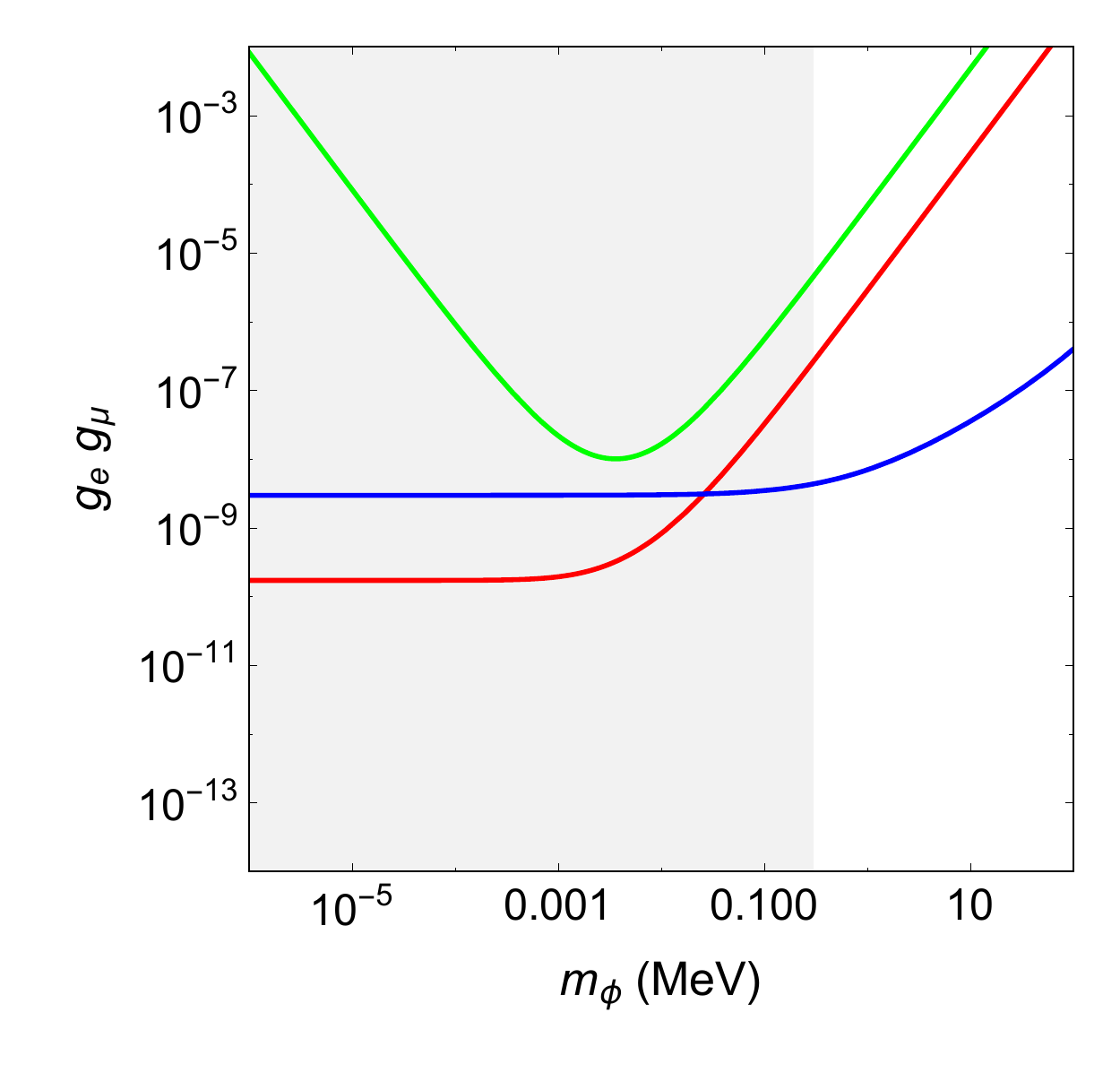}
\caption{Constraint on the dimensionless coupling  $g_e \times g_{\mu}$ as a function of the scalar/vector mass. The blue curve represents the bound coming from  the product of the measurement of the electron gyromagnetic factor $a_e$ \cite{g2e,giudice2} and the muonic ($ 5 \sigma$ bound) $a_{\mu}$ \cite{g2m}, while the red curve is the current bound extracted by Mu $1S- 2S$ transition, Eqs.~(\ref{Mu1s2sexp}) and (\ref{Mu1s2stheo}). The green curve corresponds to the current sensitivity of the Lamb Shift measurement \cite{Woodle:1990ky}.}
\label{muonium1}
\end{center}
\end{figure}
 \begin{figure}[t]
\begin{center}
\includegraphics[scale=0.6]{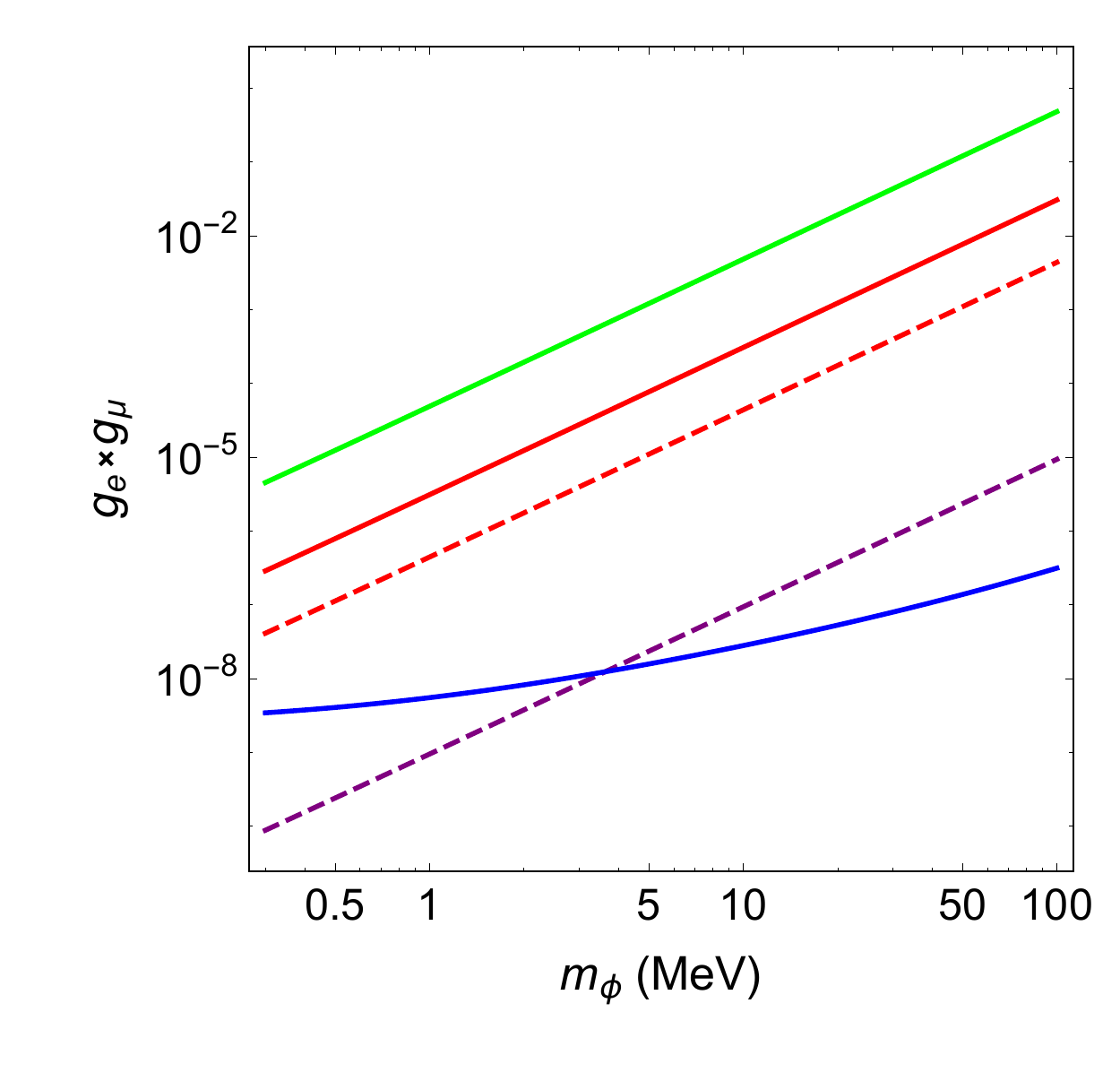}
\caption{Constraint on the dimensionless coupling  $g_e \times g_{\mu}$ as a function of the scalar/vector mass. As in Fig.~\ref{muonium1}, the blue curve represents the bound coming from  the product of the measurement of the electron gyromagnetic factor $a_e$ \cite{g2e,giudice2} and the muonic  $a_{\mu}$ \cite{g2m} while the red curve is the current bound extracted by Ps $1S-2S$ transition \cite{helium,Ps1s2s}. The green curve corresponds to the current sensitivity of the Lamb Shift measurement \cite{Woodle:1990ky}.
The dashed red curve is the $1S-2S$ projected sensitivity assuming that the experimental precision will match the theoretical one  \cite{PhysRevA.59.4316}. The dashed purple is the $1S-2S$ sensitivity considering an improvement of the theoretical and experimental error (Mu-MASS \cite{Crivelli:2018vfe}) down to 3 kHz. This would require an improvement of the muon mass measurement like the one planned at  MUSEUM (J-PARC) \cite{Crivelli:2018vfe,MUSEUM}. }
\label{muonium2}
\end{center}
\end{figure}

\paragraph{Lamb Shift}

The theoretical prediction for the Lamb shift in muonium can be obtained from the expressions in \cite{Peset:2015zga,Eides:2000xc}. It reads
\be
(E(2S_{1/2})-E(2 P_{1/2}))^{\text{th}} _{\rm Mu} = 1047.284(2)  \; \rm MHz.
\ee
In this case, the error is in fact dominated by the QED computation and estimated by the $\mathcal{O}(m_\mu\alpha^8\ln^3\alpha)$ contribution. 
The best experimental neasurement at the moment \cite{Woodle:1990ky} is 
\be
(E(2S_{1/2})-E(2 P_{1/2}))^{\text{exp}} _{\rm Mu} = 1042(22)  \; \rm MHz.
\ee
Its large uncertainty is the biggest limit to reach to new physics. 

Fig.~\ref{muonium1} shows the sensitivity to new physics of the state-of-the-art precise Mu spectroscopy. 
In the massless limit the Mu bound is an order of magnitude stronger than the product of the two gyromagnetic factors (even though a 5$\sigma$ bound is taken here to account for the current tension in the value of $a_\mu$). However, as discussed in the previous section, the electron coupling is constrained by astrophysics for mediators lighter than 300 keV, while the Mu constraint reads as:
\begin{equation}
g_e \times g_{\mu} \lesssim 10^{-10} \times \frac{\Delta}{9.8 \; \rm MHz} \end{equation}
where $\Delta $ is the experimental/theoretical error. It is thus clear that it would be extremely challenging to compete with Eq.~(\ref{astro}). For this reason, Fig.~\ref{muonium2} focuses on the heavy mass region showing that even a modest improvement of the experimental precision to match the current theoretical precision could deliver interesting results.

\section{Spin-dependent forces}

Spin-dependent forces could arise either via a pseudo-scalar or pseudo-vector mediator. In the following we will focus on the first case, while most of the literature \cite{KNP1,KNP2} studied the latter where the contribution to atomic observables such as the hyperfine splitting (HFS) is larger. 
Let us notice that the sensitivity of spectroscopy to light axion-like particles is very limited compared, for instance, to the one to spin-independent new forces, differently to the $a_e$ case.
Previous attempts~\cite{Ozeri} in the literature focus only on the very low mass region ($ m_{\phi} < 1/a_{0,e}$) where, however, similarly to the case of spin-independent forces,  the possibility of a BSM discovery is unlikely and limited to quite exotic scenarios due to the strong bounds from astrophysics, $g_{e}^{\rm ALP} \lesssim 10^{-13}$ for $ m_{\rm ALP} \lesssim 10$ keV \cite{Raffelt:2012sp}.
\par
The existence of a novel massive pseudoscalar field interacting with electrons and muons would lead to a lepton-antilepton interaction that reads \cite{bogdan,newbudker,Claudia2018}:

\begin{eqnarray}
V_{\text{ALP}}(\mathbf{r})=-\frac{g_{i}^{\rm ALP} g_{j}^{\rm ALP}}{12 \pi m_i m_j } \big[ \mathbf{S}_1 \cdot \mathbf{S}_2 \left(4\pi\delta^3(\mathbf{r})-\frac{m_{\text{ALP}}^2}{r} \right)- \nonumber \\
\frac{S_{12}(\hat{\mathbf{r}})}{4}\left(\frac{m_{\text{ALP}}^2}{ r}+\frac{3}{r^3}+\frac{3 m_{\text{ALP}}}{ r^2} \right) \big]e^{-r m_{\text{ALP}}}.\nonumber \\
\label{ALP}
\end{eqnarray}

\noindent
where $g_{i,j}^{\rm ALP}$ is the dimensionless coupling constant to the lepton $i,j$ and $m_{\rm ALP} $ the new particle's  mass. $\mathbf{S}_i$ is the spin of the i-th lepton and $m_i$ its mass. The tensor operator $S_{12}(\hat{\mathbf{r}})=4[3(\mathbf{S}_1\cdot \hat{\mathbf{r}})(\mathbf{S}_2\cdot \hat{\mathbf{r}})-\mathbf{S}_1 \cdot \mathbf{S}_2]$ is only relevant when $l\neq 0$. 

The energy levels for general quantum numbers produced by the potential given in Eq.~(\ref{ALP}) can be found in \cite{Claudia2018}. They produce spin-dependent as well as spin-independent energy shifts of the different states of Ps and Mu. This can be seen from the decomposition in terms of the total spin of the bound state of the operator $\mathbf{S}_1 \cdot \mathbf{S}_2=1/2\mathbf{S}^2-3/4$. Therefore it will contribute to the hyperfine splitting (HFS) of the different states as well as to the $1S-2S$ transition.

\subsection{ Positronium}

\paragraph{Ground state HFS}

The HFS on the Ps ground state, $\Delta\nu=E(1^3S_1)-E(1^1S_0)$, has been extensively studied, both theoretically and experimentally. In particular, its experimental measurement has been dramatically improved up to the few MHz of precision level~\cite{PhysRevA.27.262,PhysRevA.30.1331,Ishida:2013waa}.  The latest and most precise measurement leads to \cite{Ishida:2013waa}
\be
(\Delta \nu)^{\text{exp}} _{\text{Ps}} =203394.2(2.1)  \;  \rm MHz.
\ee

The theoretical computation is completely known up to $\mathcal{O}(m_e\alpha^6)$ \cite{PhysRevA.59.4316} together with the leading logarithmic $\mathcal{O}(m_e\alpha^7 \ln\alpha)$ and $\mathcal{O}(m_e\alpha^7 \ln^2\alpha)$ corrections  \cite{Karshenboim:1993,Hill:2000zy,PhysRevLett.86.1498,Kniehl:2000cx}. The computation of the finite piece of $\mathcal{O}(m_e\alpha^7)$ remains subject to much theoretical interest \cite{Adkins:2018lvj} but is still incomplete. We quote here the value obtained in \cite{Baker:2014sua} where the a priori largest effects at such order have been added

\be
(\Delta \nu)^{\text{th}} _{\text{Ps}}= 203391.91(22) \;  \rm MHz.
\ee

 The measurements of $(\Delta \nu)_{\text{Ps}}$ are still far from the theoretical prediction. Indeed  the \cite{PhysRevA.27.262,PhysRevA.30.1331} measurements exhibit a $\sim 3\sigma$ discrepancy with the theoretical calculation, while the latest \cite{Ishida:2013waa} is in agreement with it, disfavoring the previous ones. 
A new direct measurement is planned at ETHZ \cite{crivelliPSHFS} which will reach a precision of 10ppm.

\begin{figure}[h]
\begin{center}
\includegraphics[scale=0.6]{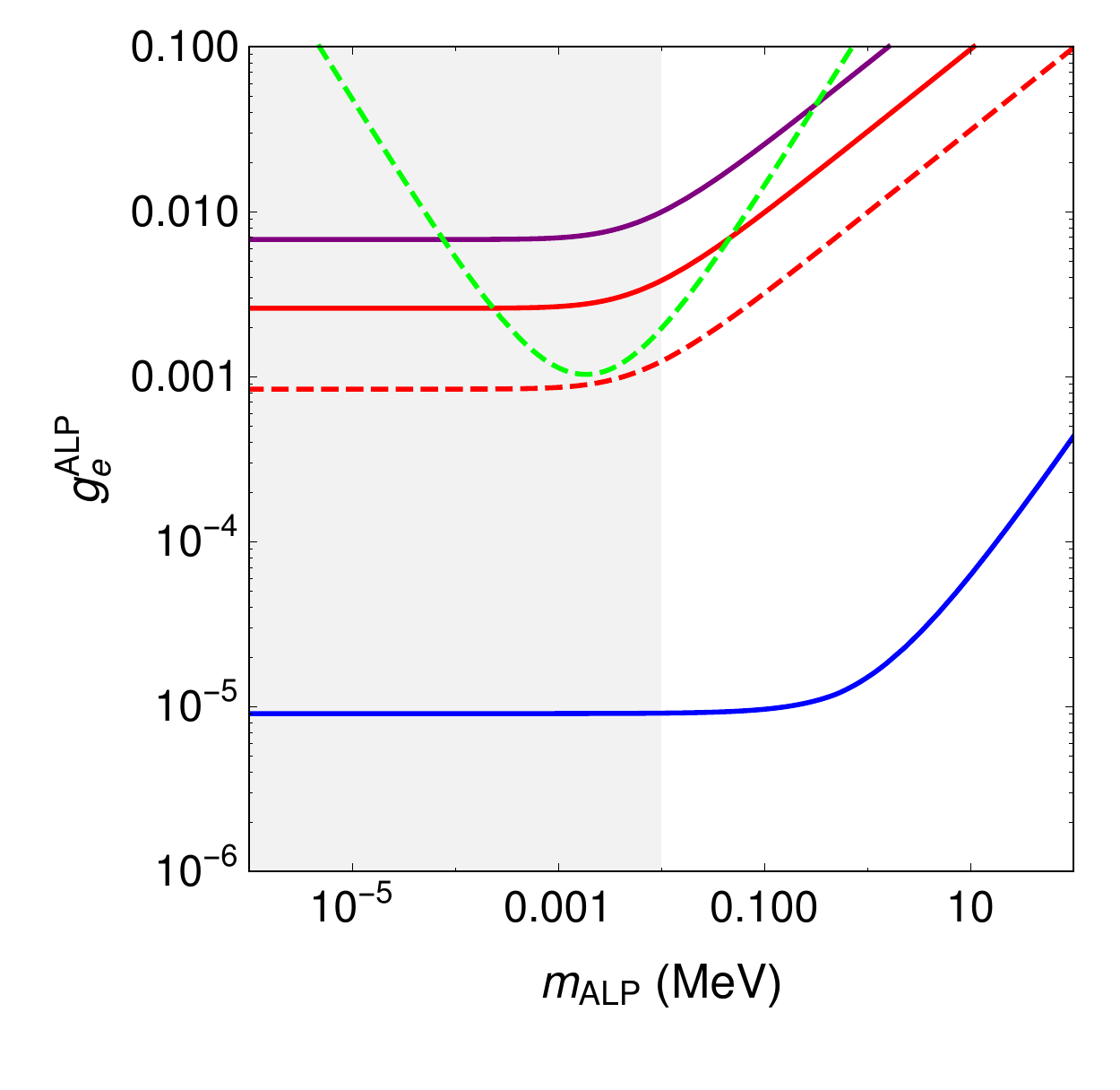}

\caption{Constraint on the spin-dependent dimensionless coupling $g_{e}^{\rm ALP}$ as a function of the axion mass. The blue curve represents the bound coming from the measurement of the electron gyromagnetic factor $a_e$ \cite{g2e,giudice2}, while the red the current Ps sensitivity considering the MHz level latest measurement  \cite{Ishida:2013waa}.
The limit from the current $1S-2S$ measurement \cite{Ps1s2s} corresponds to the purple line. The gray region is excluded by stellar cooling constraints, that $g_{e}^{\rm ALP} \lesssim 10^{-13}$ for $ m_{\rm ALP} \lesssim 10$ keV \cite{Raffelt:2012sp}. The dashed red curve corresponds to a projected sensitivity assuming that the future experimental precision for the ground state HFS would match the theoretical prediction \cite{Baker:2014sua}. The dashed green curve is the projection for the ultrafine splitting assuming experimental precision to match the theoretical one \cite{Lamm}.}
\label{fig4}
\end{center}
\end{figure}

Fig.~\ref{fig4} shows the existing and projected constraints on the dimensionless coupling constant $g_{e}^{\rm ALP} $ as a function of the ALP mass. The red curve corresponds to the bound from the HFS of the ground state Ps (the dashed curve is the projection assuming the experimental precision will match the theoretical one). We notice that also in this case the Ps sensitivity is several orders of magnitude suppressed compared to the gyromagnetic factor $a_e$ even in the massless limit.
In order to improve the sensitivity in the heavy mass region, $m_{\rm ALP} > 10$ keV, an at present unrealistic improvement of seven orders of magnitude would be necessary to match the current sensitivity of $a_e$.
 \paragraph{Ultrafine splitting}
Here we consider the ultrafine splitting between the single $2^1P_1$ state and the spin-average of the triplet $2^3P_J$:
\be
\Delta_{2,P}\equiv E(2 ^1P_1)-\frac{1}{9}\left(E(2 ^3P_0)+3E(2 ^3P_1)+5E(2 ^3P_2) \right).
\ee
 Only the operator $\mathbf{S}_1 \cdot \mathbf{S}_2$ in Eq.~(\ref{ALP}) contributes to this splitting. $\Delta_{2,P}$ has been calculated to leading $\mathcal{O}(m_e\alpha^6)$ accuracy in QED~\cite{PhysRevA.59.4316}, leading to 

\be
\Delta_{2,P}=\frac{683 m_e\alpha^6}{172800}=73.7(2.6) \;  \rm kHz,\label{eq:UFS}
\ee

\noindent
which is almost two orders of magnitude more precise than the current experimental observations~\cite{Ley1994,PhysRevLett.71.2887}. 

The green curve in Fig.~\ref{fig4} is the bound based on the ultrafine splitting in Eq.~(\ref{eq:UFS}) assuming a measurement able to reach a precision comparable to the theoretical error.

\paragraph{$1S-2S$ transition} The difference between Eqs.~(\ref{PsLS}) and~(\ref{PsLSexp}) can also be produced by the potential in Eq.~(\ref{ALP}). This bound is typically not taken into account for a pseudoscalar mediator since it is $\alpha^2$ suppressed in the massless limit compared to the scalar, but, as one can see from the purple line in Fig.~\ref{fig4}, its bound is in fact competitive with the one of the HFS. This is specially so for large axion masses where the contribution grows like $\sim \frac{m_{\rm ALP^2}}{m_e^2}$.


\paragraph{Other splittings}

The theoretical QED computations in \cite{PhysRevA.59.4316} together with the experimental results in \cite{Ley1994,PhysRevLett.71.2887} and the computations for axions and scalars in \cite{Claudia2018}, allow us to obtain bounds from other energy splittings such as the fine splittings or weighted combinations $S$- and $P$-wave states. We find from these splittings very similar bounds to those shown in Fig.~\ref{fig4}, and so we don't explicitly show them here.

\subsection{Muoniun}
\begin{figure}[t]
\begin{center}
\includegraphics[scale=0.6]{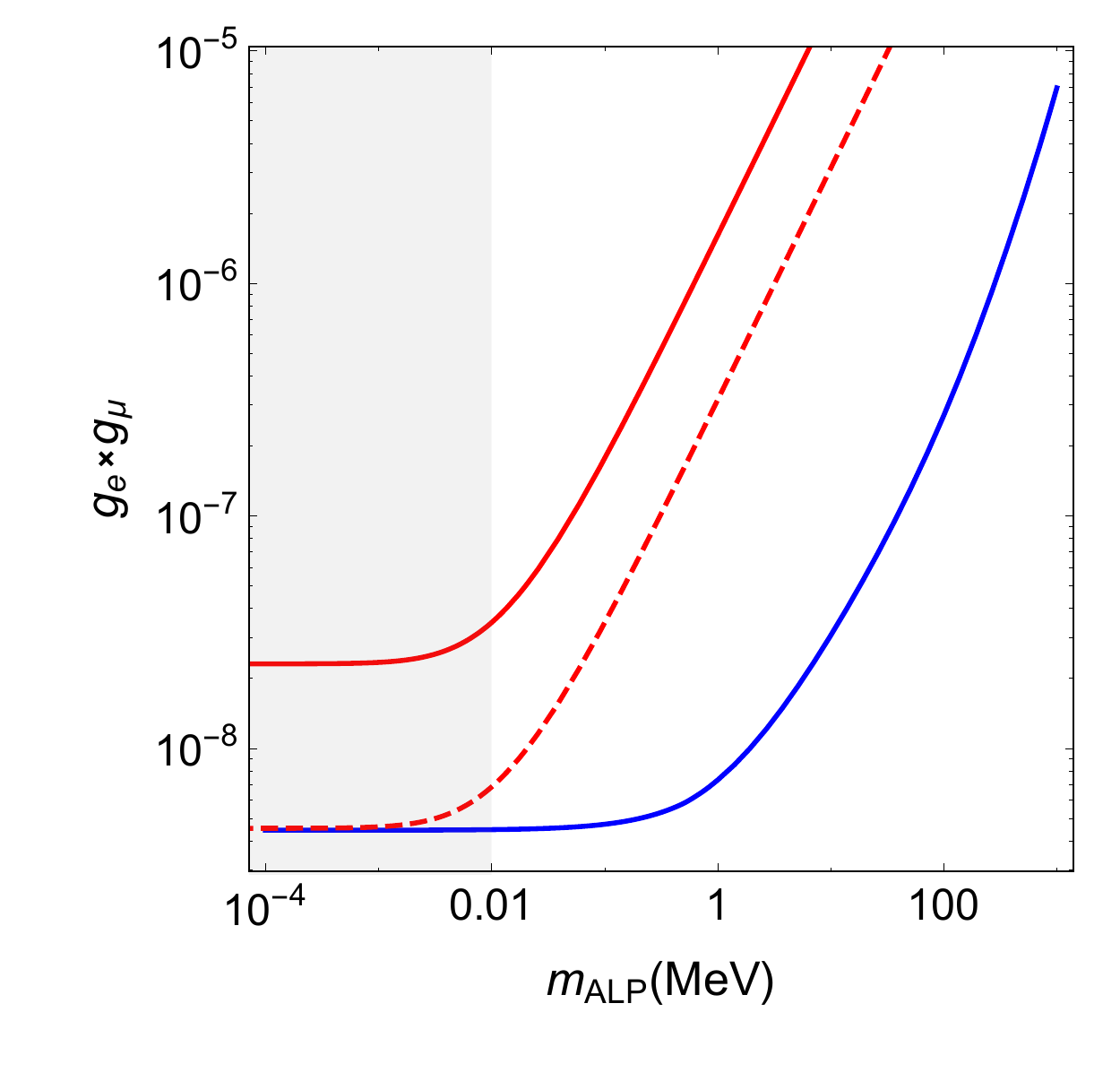}
\caption{Constraint on the spin-dependent dimensionless coupling $g_{e}^{\rm ALP} \times g_{\mu}^{\rm ALP} $ as a function of the axion mass. The blue curve corresponds to the product of  the bound arising from the electronic and ($2 \sigma$) muonic gyromagnetic factors $a_e, a_{\mu}$ \cite{g2e,giudice2,g2m}, while the red  (dashed red) one is the constraint from the muonium HFS limited by the  theoretical (experimental) precision \cite{Karshenboim20051,mckeenK}.
The gray shaded limit is the bound obtained combining the astrophysical one on the electron coupling \cite{Raffelt:2012sp} and the $5 \sigma$ $a_{\mu}$ \cite{g2m}. }
\label{muoniumSD}
\end{center}
\end{figure}

The latest experimental measurement of the muonium ground state HFS was performed at Los Alamos Meson Physics Facility (LAMPF) and it gives \cite{osti_295569}:
\be
(\Delta \nu)^{\text{exp}} _{\rm Mu}=4463 302.765 (53) \; \rm kHz.\label{HFSMuexp}
\ee
In the near future the MuSEUM project at JPARC is planning to perform a new measurement to improve the LAMPF result  thanks to a new and intense beam line  \cite{Crivelli:2018vfe,MUSEUM}.

For the ground state HFS in Mu the current theoretical prediction is \cite{Karshenboim20051,mckeenK}:
\be
(\Delta \nu)^{\text{th}} _{\rm Mu}=4463 302.89 (27) \; \rm kHz
\ee
so the experimental precision is slightly better than the theoretical one. The latter is limited by the uncertainty of the muon mass measurement. As for the $1S-2S$ analysis, we use here the muon mass measurement from the Breit-Rabi splittings. This measurement and the experimental value in Eq.~(\ref{HFSMuexp}) have a small correlation that is negligible for our purpose, as are the possible effects of new physics in the determination of the muon mass.

\par
In Fig.~\ref{muoniumSD} we see that the sensitivity for  current and future measurements in the large mass region cannot compete with the existing $g-2$ constraints. 
The ALP-lepton coupling leads to a negative contribution to the leptonic gyromagnetic factor which cannot explain the $a_{\mu}$ anomaly and is instead strongly constrained by it. Therefore, in this case, we considered excluded the region $a_{\mu} >2 \sigma$.



\section{Conclusions and prospects}
We studied the sensitivity of Ps and Mu spectroscopy to spin-dependent or -independent dark forces between electrons and muons. 
Our main findings are the following:
\begin{itemize}
\item  The sensitivity of the electronic gyromagnetic of the electron to dark sector fields is significantly stronger than Ps spectroscopy, even for Rydberg states. Hence, these measurements could be considered new physics free unless the experimental precision is reduced down to Hz level, which is unrealistic for the near future. 
However, this  provides a motivation to apply modern spectroscopic techniques such as two-photon transitions, electromagnetic induced transparency approaches through a dark state and other techniques borrowed from modern quantum optics.

\item  Mu precision spectroscopy has a more interesting potential to probe new physics.  Future measurements planned at PSI and J-PARC have the potential to set world record bounds for medium range spin independent interactions (shorter than 5  MeV) between electrons and muons. 
\end{itemize}
 \section{Acknowledgements}
We thank Prof. D. Cassidy and Dr. P. Crivelli for providing important information on the precision of the upcoming experiments and Dr. S. Karshenboim for useful discussions. CP acknowledges the financial support from Fundaci\'on Ram\'on Areces.

\bibliography{lightatoms-bib}

\end{document}